\documentclass{ws-procs975x65}

\begin{document}



\title{QUASI-STATIONARY ROUTES TO THE KERR BLACK HOLE}

\author{REINHARD MEINEL}

\address{University of Jena, Theoretisch-Physikalisches Institut,\\
Max-Wien-Platz 1, 07743 Jena, Germany\\
\email{meinel@tpi.uni-jena.de}}

\begin{abstract}
Quasi-stationary (i.e.~parametric) transitions from rotating equilibrium configurations of fluid bodies to rotating black holes are discussed. For the idealized model of a rotating disc of dust, analytical results derived by means of the ``inverse scattering method'' are available. They are generalized by numerical results for rotating fluid rings with various equations of state. It can be shown rigorously that a black hole limit of a fluid body in equilibrium occurs if and only if the gravitational mass becomes equal to twice the product of angular velocity and angular momentum. Therefore, any quasi-stationary route from fluid bodies to black holes passes through the extreme Kerr solution.
\end{abstract}

\bodymatter

\section{Introduction}
The exterior metric of a spherically symmetric star, even in the case of collapse, is always given by the Schwarzschild metric. This is a consequence of Birkhoff's theorem. Therefore, the collapse of a sufficiently massive, non-rotating star at the end of its life leads quite naturally to a Schwarzschild black hole, as in the idealized case of the Oppenheimer--Snyder dust collapse. On the other hand, a continuous quasi-static transition from stars (modelled as perfect fluid spheres) to black holes is not possible (cf.~Buchdahl's inequality). Without rotation, the black hole state can only be reached dynamically. 

For rotating stars, the situation is different in both previously mentioned respects. Firstly, the exterior metric is not the Kerr metric in general. (There is no analogue to Birkhoff's theorem in this case.) It is generally believed, based on the cosmic censorship conjecture combined with the black hole uniqueness theorems, that the collapse of a rotating star leads asymptotically to the Kerr black hole, i.e.~to the Kerr metric outside the horizon. This has not yet been proved however. But secondly, a continuous quasi-stationary transition from rotating perfect fluid bodies to rotating black holes is possible. In the following, this will be demonstrated by reviewing analytical results for a rotating disc of dust as well as numerical results for rotating fluid rings. Moreover, necessary and sufficient conditions for a black hole limit of rotating fluid bodies in equilibrium will be discussed.

\section{From rotating discs and rings to black holes}
Bardeen and Wagoner\cite{bw} solved the general relativistic problem of a uniformly rotating disc of dust approximately. They found evidence that in a certain parameter limit the solution approaches the extreme Kerr metric outside the horizon. This has been confirmed by the exact solution to the disc problem\cite{nm1,nm2,nm3} derived by means of the ``inverse scattering method''. The solution depends on two parameters, say the gravitational mass $M$ and the angular momentum $J$. Other parameters, such as the disc's angular velocity $\Omega$ (as seen from infinity\footnote{We assume asymptotic flatness.}), are then functions of $M$ and $J$.  In the black hole limit, the relations $J=M^2$ and $M=2\Omega J$ hold\footnote{We use units in which $G=c=1$.}. It should be mentioned that a ``separation of spacetimes'' occurs in the parameter limit. From the ``exterior point of view'', the extreme Kerr metric outside the horizon emerges, whereas from the ``interior point of view'' a non-asymptotically flat spacetime containing the rotating disc emerges, which approaches the extreme Kerr throat geometry (``near-horizon geometry'') at infinity. More details can be found in Refs.~\citen{bw} and \citen{m1}.

The separation of spacetimes mentioned above, which turns out to be a universal feature in the limit, allows for the existence of a black hole limit independent of the fluid body's topology. Indeed, such a limit was found numerically for bodies of toroidal topology, the ``relativistic Dyson rings''\cite{akm} and their generalizations\cite{fha,lpa}. So far, these ring solutions with various equations of state are the only known examples of genuine fluid bodies permitting a black hole limit. For a review of relativistic equilibrium configurations of constant mass-energy density --- including the relativistic Dyson rings --- see Ref.~\citen{afkmps}. 

\section{Conditions for a black hole limit}
It can be proved that the parameter relation
\begin{equation}
M=2\Omega J
\label{bhl}
\end{equation}
is necessary\cite{m2} and sufficient\cite{m3} for a (Kerr) black hole limit of rotating fluid bodies in equilibrium. This shows once again that such a limit is impossible without rotation. Moreover, since $\Omega$ must become equal to the ``angular velocity of the horizon'' of the Kerr black hole,
\begin{equation}
\Omega^{\rm H}= \frac{J}{2M^2\left[M+\sqrt{M^2-(J/M)^2}\,\right]}\, ,
\label{OmH}
\end{equation}
the relation
\begin{equation}
J=M^2,
\label{ek}
\end{equation}
characteristic of an {\it extreme\/} Kerr black hole, must hold in the limit. Therefore, any quasi-stationary route from fluid bodies to black holes passes through the extreme Kerr solution. Note that, in contrast to (\ref{bhl}), the relation (\ref{ek}) alone is not sufficient for a black hole limit of a fluid body in equilibrium. Indeed, there exist normal fluid configurations with $J<M^2$, $J=M^2$ as well as $J>M^2$ (the disc and ring solutions discussed above, however, have $J>M^2$ except for the black hole limit where $J=M^2$). But fluid configurations always satisfy $M>2\Omega J$, and $M=2\Omega J$ ($=2\Omega^{\rm H}J$) is approached precisely in the black hole limit. Non-extreme Kerr black holes (characterized by $J<M^2$) again satisfy $M>2\Omega^{\rm H} J$. 

\section{Outlook}
It is an open question, whether there are sequences of {\it stable\/} equilibrium fluid configurations approaching a black hole limit continuously, i.e.~whether quasi-stationary routes to the Kerr black hole as discussed here are to be expected in the real world. It may well be that a configuration which is already close to the black hole limit will dynamically evolve towards a slightly sub-extreme Kerr black hole as a result of small perturbations. Investigations in this direction may lead to further interesting insights concerning questions of gravitational collapse, black hole formation and cosmic censorship.

\section*{Acknowledgments}
I would like to thank Marcus Ansorg, Andreas Kleinw\"achter, Gernot Neugebauer and David Petroff for valuable discussions. This research was supported by the Deutsche Forschungsgemeinschaft (DFG) through the SFB/TR7 ``Gravitations\-wellenastronomie''.

\end{document}